\documentclass[sigconf]{acmart}
\AtBeginDocument{%
  }

\usepackage{booktabs}
\usepackage{multirow}
\usepackage{graphicx} 
\usepackage{pifont}

\usepackage{array}   
\usepackage{balance}
\usepackage{natbib}

\usepackage{booktabs}
\usepackage{siunitx}
\usepackage{booktabs}

\usepackage{subcaption}
\usepackage{float}

\newcommand{\cmark}{\checkmark}
\newcommand{\xmark}{\times}
\renewcommand{\arraystretch}{1.35}

\copyrightyear{2026}
\acmYear{2026}
\setcopyright{cc}
\setcctype{by}
\acmConference[SIGIR '26]{Proceedings of the 49th International ACM SIGIR Conference on Research and Development in Information Retrieval}{July 20--24, 2026}{Melbourne, VIC, Australia}
\acmBooktitle{Proceedings of the 49th International ACM SIGIR Conference on Research and Development in Information Retrieval (SIGIR '26), July 20--24, 2026, Melbourne, VIC, Australia}
\acmDOI{10.1145/3805712.3808551}
\acmISBN{979-8-4007-2599-9/2026/07}

\begin{document}

\title{A Reproducibility Study of \\Metacognitive Retrieval-Augmented Generation}

\author{Gabriel Iturra Bocaz}
\email{gabriel.e.iturrabocaz@uis.no}
\orcid{0009-0001-9635-0683}
\affiliation{%
  \institution{University of Stavanger}
  \city{Stavanger}
  \country{Norway}
}

\author{Petra Galu\v{s}\v{c}\'{a}kov\'{a}}
\email{petra.galuscakova@uis.no}
\orcid{0000-0001-6328-7131}
\affiliation{%
  \institution{University of Stavanger}
  \city{Stavanger}
  \country{Norway}
}

\renewcommand{\shortauthors}{Gabriel Iturra Bocaz and Petra Galuščáková}

\begin{abstract}
  Recently, Retrieval Augmented Generation (RAG) has shifted focus to multi-retrieval approaches to tackle complex tasks such as multi-hop question answering. However, these systems struggle to decide when to stop searching once enough information has been gathered. To address this, \citet{zhou2024metacognitive} introduced Metacognitive Retrieval Augmented Generation (MetaRAG), a framework inspired by metacognition that enables Large Language Models to critique and refine their reasoning. In this reproducibility paper, we reproduce MetaRAG following its original experimental setup and extend it in two directions: (i) by evaluating the effect of PointWise and ListWise rerankers, and (ii) by comparing with SIM-RAG, which employs a lightweight critic model to stop retrieval. Our results confirm MetaRAG's relative improvements over standard RAG and reasoning-based baselines, but also reveal lower absolute scores than reported, reflecting challenges with closed-source LLM updates, missing implementation details, and unreleased prompts. We show that MetaRAG is partially reproduced, gains substantially from reranking, and is more robust than SIM-RAG when extended with additional retrieval features.
\end{abstract}




\maketitle

\section{Introduction}

Metacognition is the ability to reflect on and critique one’s own cognitive processes \cite{lai2011metacognition}. It enables individuals to notice gaps in understanding, judge whether additional information is needed, and adjust their strategies to reach more reliable conclusions. This self-awareness is fundamental in tasks that demand several reasoning steps across multiple pieces of information, allowing errors to be detected and planning strategies to be refined \cite{schraw1995metacognitive,lai2011metacognition}. Such ability is important because it helps to maintain reasoning that stays focused on the question rather than drifting to irrelevant details, and that remains consistent when integrating new or conflicting information \cite{gotoh2016development}. However, although modern LLMs employ implicit control mechanisms such as routing or model selection, they typically generate responses without explicit self-reflection or awareness of their own knowledge limitations \cite{li2024hindsight,shinn2023reflexion}.

Inspired by the concept of metacognition, \citet{zhou2024metacognitive} proposed MetaRAG, a framework that enables RAG systems to monitor and critique their own reasoning before  further retrieval is required. While other reasoning-oriented approaches such as CoT \cite{wei2022chain}, ReAct \cite{yao2023react}, or Self-Ask \cite{press2022measuring} focus on guiding the reasoning process, they do not explicitly implement metacognitive control. This distinction remains important because multi-stage retrieval systems must regulate additional retrieval to avoid unnecessary overhead, latency, and cost, especially in large-scale and industrial settings. In this reproducibility paper, we reproduce MetaRAG's results under a comparable setup and further investigate how aspects such as document ordering relate to metacognitive reasoning in LLMs through reranker models.

In this work, we formulate and address the following research questions:

- \textbf{RQ1}: How reproducible e are the results of MetaRAG?

While the original MetaRAG paper reports strong results on multi-hop Question-Answering (QA) tasks \cite{mavi2024multi}, its reproducibility remains uncertain \cite{laskar2024systematic}. For example, the code and prompts for the baseline implementations are not fully available, and the underlying LLMs have evolved since the study was published\footnote{\url{https://platform.openai.com/docs/deprecations\#2023-11-06-chat-model-updates}}. Moreover, some aspects of the original setup remain unclear, such as how sparse and dense retrieval are combined in the implementation and described in the paper \cite{zhou2024metacognitive}. Because MetaRAG represents an important step toward integrating metacognitive control into RAG systems, we aim to re-examine how it performs under current experimental conditions, including updated LLM backends, retrieval implementations, and incomplete or underspecified setup details.


- \textbf{RQ2}: Can reranker models improve the performance of MetaRAG?

In the human reasoning process for answering complex questions, we first prioritize information and then reflect on it \cite{nelson1990metamemory}. Similarly, while MetaRAG critiques its own reasoning, it does not explicitly address the quality or ordering of retrieved documents. Since reranker models help reduce noise and control order~\cite{glass2022re2g,yu2024rankrag,moreira2024enhancing}, we investigate whether integrating them into MetaRAG enhances its overall reasoning and answer quality.

- \textbf{RQ3}: How does MetaRAG compare to other metacognitive RAG frameworks, such as SIM-RAG?

While MetaRAG investigates metacognition in multi-hop QA, the comparative impact of different metacognitive mechanisms on retrieval and reasoning is still not well understood. For example, \citet{yang2025knowing} argue that multi-round retrieval RAG systems do not know when they have gathered enough information to answer a question. They propose SIM-RAG, a framework that adds a lightweight critic module which continuously checks if enough information has been retrieved, helping the system decide when to stop searching and start reasoning. We therefore compare MetaRAG with SIM-RAG to better understand how different forms of metacognition control influence retrieval stopping and reasoning performance.

\textbf{Main Contributions.} Our main contributions are summarized as follows: (1) we conduct a reproducibility study of the MetaRAG framework, focusing on evaluating its performance on QA tasks and comparing it with other baselines, (2) we extend MetaRAG with rerankers, finding performance gains by reducing noise and mitigating the effects of document reordering, and (3) we compare MetaRAG with SIM-RAG and show that MetaRAG, especially when combined with rerankers, performs more robustly. To support further research, we release our code\footnote{\url{https://github.com/iai-group/sigir2026-metarag}} and experimental setup, including baselines.

\section{Related Work}

RAG has become a central paradigm for enhancing LLMs by injecting external information, which extends their internal knowledge and addresses common issues such as hallucinations and unverifiable answers \cite{lewis2020retrieval,izacard2021leveraging}. Early RAG systems employed single-retrieval strategies \cite{he2021efficient,izacard2021leveraging,lewis2020retrieval,zhong2022training} to answer simple queries, such as factual questions, but these approaches fall short for multi-hop QA tasks that require several reasoning steps to produce high-quality answers. To overcome these challenges, research has shifted towards multi-retrieval frameworks \cite{khandelwalgeneralization,trivedi2022interleaving,khotdecomposed,yao2023react}, in which the retrieval stage is invoked iteratively during reasoning or generation. Generally, these approaches can be broadly divided into passive retrieval, decompositional calls, and dynamic schemes controlled by LLMs.

\textit{Passive Retrieval}. In these approaches, retrieval is triggered on a fixed schedule (e.g., after every sentence or token count), regardless of whether the system needs to retrieve information at that moment. For example, \citet{khandelwalgeneralization} enhance traditional language models by consulting a datastore of past contexts to recall similar contexts for informing the next predictions. Similarly, \citet{trivedi2022interleaving} introduced IR-CoT, which performs retrieval after every Chain of Thought (CoT) reasoning step.

\textit{Decompositional Queries}. In these systems, complex queries are divided into sub-questions, each triggering a dedicated retrieval process, the results of which are later aggregated. \citet{khotdecomposed} proposed that rather than solving a complex problem all at once, the original question should be decomposed into a sequence of simpler sub-queries, with each sub-question performing retrieval independently.

\textit{Retrieval controlled by LLMs}. In these frameworks, retrieval is guided by the reasoning steps of LLMs. For example, ReAct \cite{yao2023react} combines CoT reasoning to develop a series of thoughts that influence actions in an environment, including retrieval, to obtain more information at each specific step in the reasoning process. Furthermore, the synergy of these ``reasoning'' frameworks and iterative retrieval has enabled RAG systems to address more complex tasks, such as multi-hop QA.

Recent work on multi-hop QA has focused on improving reliability and control in RAG particularly in the presence of noisy retrieval and over-searching. Prompt-based strategies such as Self-Consistency~\cite{wang2022selfconsistency} and Context Repetition (CoRe)~\cite{yu2025unleashing}, as well as fine-tuning approaches with external critics, aim to guide the reasoning process beyond a single retrieve–generate step. Among these approaches, SIM-RAG~\cite{yang2025knowing} introduces a lightweight classifier trained on synthetic data to determine whether additional retrieval is required. This design enables adaptive retrieval decisions but provides only an implicit form of self-monitoring, as the underlying reasoning errors or evidence deficiencies are not explicitly diagnosed.

MetaRAG~\cite{zhou2024metacognitive} explicitly models metacognition through a structured \textit{monitor–evaluate–plan} loop, allowing the system to identify reasoning issues, assess evidence sufficiency, and decide whether to continue retrieval. This explicit formulation places MetaRAG within a broader line of agentic and multi-round retrieval frameworks~\cite{yao2023react,trivedi2023interleaving,asai2024self,shinn2023reflexion}, and makes it a particularly relevant case for studying the reproducibility of metacognitive control mechanisms under modern IR systems.



\section{Metacognitive RAG Frameworks}

To contextualize our study, we first describe the general structure of metacognitive RAG systems \cite{zhou2024metacognitive}. These frameworks extend the standard \textit{retrieve-then-generate} \cite{lewis2020retrieval} paradigm by introducing a metacognitive loop that enables the model to reflect on its own reasoning process before finalizing an answer.

\begin{figure*}[t]
    \centering
    \includegraphics[width=0.6\textwidth]{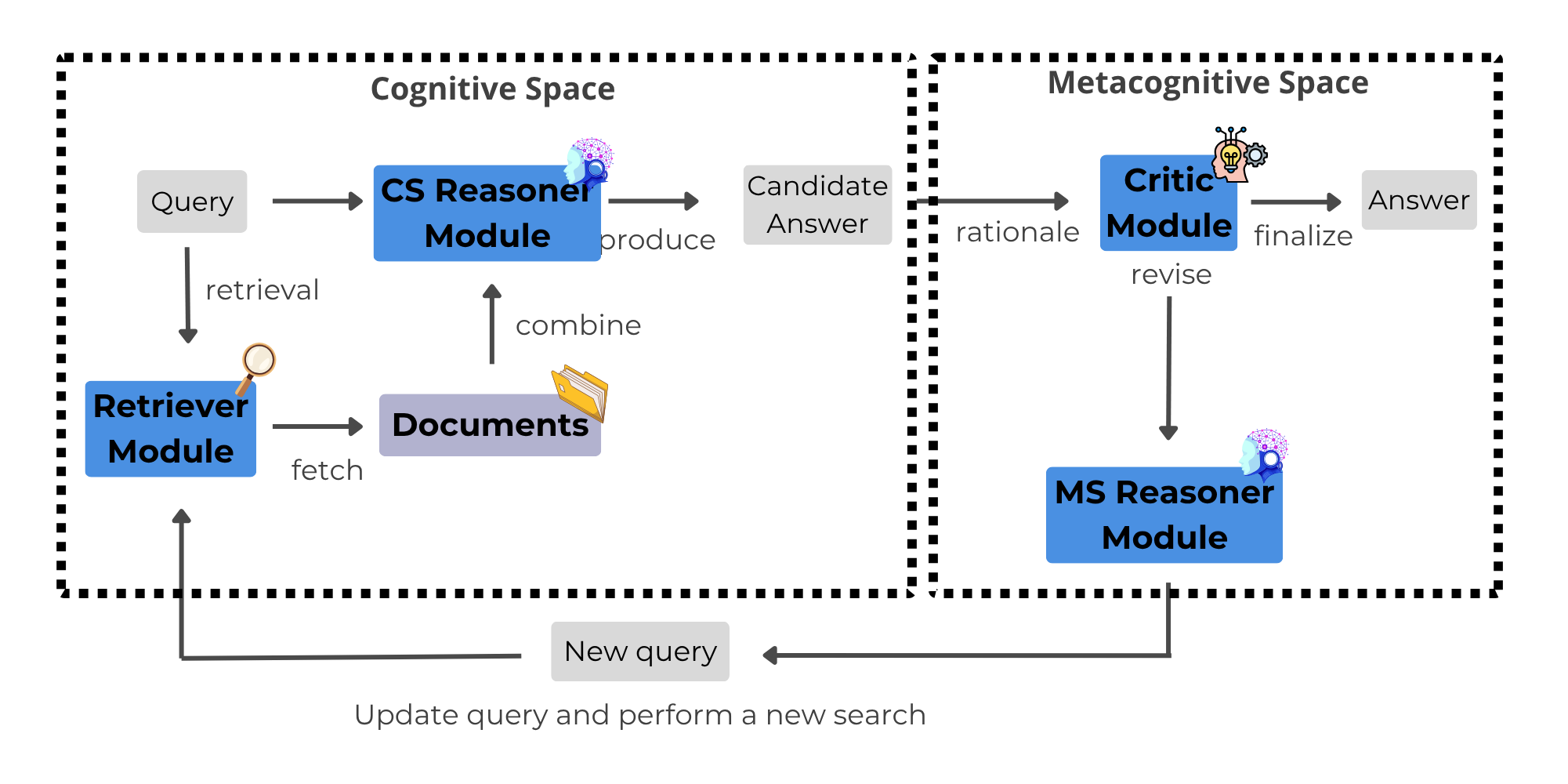}
        \caption{General architecture of metacognitive RAG systems. 
    In the \textbf{Cognitive Space (CS)}, the \textit{Retriever Module} fetches relevant documents based on the question, which are then integrated by the \textit{CS Reasoner Module} to produce a candidate answer. 
    In the \textbf{Metacognitive Space (MS)}, the \textit{Critic Module} evaluates this rationale and, if necessary, delegates to the \textit{MS Reasoner Module} to refine the reasoning or reformulate a new question for additional retrieval.}
    \label{fig:metacognitive_rag}
\end{figure*}

Figure~\ref{fig:metacognitive_rag} illustrates this general metacognitive architecture, which operates across two complementary spaces:

\begin{itemize}
    \item \textbf{Cognitive Space (CS).} This space corresponds to the standard RAG workflow. A \textit{Retriever Module} fetches relevant documents from an external knowledge base given a question, which are then combined with the question and passed to \textit{CS Reasoner Module}. This module performs the initial reasoning step, interpreting the evidence and generating a candidate answer along with its underlying rationale.
    
    \item \textbf{Metacognitive Space (MS).} In this space, a \textit{Critic Module} examines the candidate answer produced by \textit{CS Reasoner Module}, assessing its coherence, confidence, and completeness. If the rationale is insufficient, the critic invokes \textit{MS Reasoner Module} to refine the reasoning process, which may involve rewriting the question, requesting additional retrieval, or revising the generated answer before finalization.
\end{itemize}

This general pipeline serves as the foundation for the specific frameworks analyzed in this study.

\subsection{MetaRAG}

MetaRAG implements metacognitive control through an explicit \textit{monitor--evaluate--plan} loop in the MS, which determines whether the current answer should be accepted, refined, or supported by additional retrieval.

\begin{itemize}
    \item \textbf{Monitoring.}
    Given a question $q$ and retrieved passages $D_q$, the CS produces an answer $y$, while an ``expert'' QA model generates a reference answer $y'$. A similarity score between $y$ and $y'$ is computed, and metacognitive evaluation is triggered only when this score falls below a predefined threshold; otherwise, the answer is accepted directly.

    \item \textbf{Evaluation.}
    When it is activated, the Critic module diagnoses why the current answer may be unreliable. Internal knowledge sufficiency is assessed by prompting the LLM, while external knowledge sufficiency is evaluated using an NLI model that verifies whether the retrieved passages entail the required information. In addition, the Critic Module checks for common reasoning errors such as incomplete or inconsistent use of evidence.

    \item \textbf{Planning.}
    Based on the evaluation outcome, MetaRAG selects the next action. If knowledge is insufficient, the system reformulates the question by explicitly targeting missing information and retrieves additional documents. If knowledge is sufficient but the reasoning is flawed, the system revises unsupported reasoning steps or adjusts the reliance on internal versus external knowledge. This loop repeats until the answer is accepted or a maximum number of iterations is reached.
\end{itemize}

\subsection{SIM-RAG}

Instead of employing an explicit monitor--evaluate--plan loop as in MetaRAG, SIM-RAG \cite{yang2025knowing} introduces a lightweight critic module that continuously judges whether enough evidence has been retrieved to answer a question. 
This critic is trained to signal when retrieval should stop, thereby preventing unnecessary or noisy retrieval rounds. 
While MetaRAG emphasizes explicit self-reflection and reasoning refinement, SIM-RAG focuses on efficient retrieval control through implicit metacognitive signals. 
In this study, we do not aim to reproduce SIM-RAG, but include it as a comparative metacognitive framework to better contextualize the strengths and weaknesses of MetaRAG under similar experimental conditions.

\section{Experimental setting}

\subsection{Datasets and Evaluation Metrics}

Following the original paper, we conduct experiments on two multi-hop QA collections: HotpotQA \cite{yang2018hotpotqa} and 2WikiMultiHopQA \cite{tang2024multihop}. HotpotQA contains over 100K Wikipedia-based questions that often require reasoning across two or more passages, while 2WikiMultiHopQA provides 193K questions explicitly designed to enforce multi-hop reasoning across multiple documents. Both datasets are open-domain and rely on the Wikipedia corpus. 

To remain consistent with the original MetaRAG configuration, we evaluate all methods on 500 examples sampled from the development (dev) set of each collection. These samples are publicly available in our repository\footnote{\url{https://github.com/iai-group/sigir2026-metarag/tree/main/data}}. This evaluation protocol follows common practice in prior RAG works \cite{jiang2023active,yang2025knowing,zhou2024metacognitive,jeong2024adaptive,li2024smart}, where full-dataset evaluation is often infeasible due to computational constraints. Since the original MetaRAG code does not release the exact development subsets used, we randomly sample 500 examples from each dev set. We further adopt the same evaluation metrics as the original paper to assess the performance of MetaRAG and all other frameworks in this paper. At the answer level, we use exact match~(EM)~\cite{bahak2023evaluating}, which counts a prediction as correct only if it matches the gold answer string exactly, with no extra or missing words. At the token level, we use token-level F1, Precision (Prec.), and Recall (Rec.).

\subsection{Baselines and SIM-RAG}

The original MetaRAG paper does not release the code or prompts for its baseline methods, nor does it describe how these baselines are adapted for a fair comparison with MetaRAG. In particular, it remains unclear how each reasoning frameworks \cite{schulhoff2024prompt} is integrated with the retrieval component. For example, ReAct \cite{yao2023react} interleaves reasoning steps with retrieval actions, but the MetaRAG paper does not specify how this action interface interacts with the index or how it is managed across iterations. Similarly, the prompting setup used in these baselines is not documented, even though performance in reasoning-oriented methods strongly depends on prompt design~\cite{schulhoff2024prompt}. To ensure a consistent and transparent comparison, we re-implement all baselines in LangChain\footnote{\url{https://www.langchain.com/}}, explicitly defining how each method interacts with the retriever and adapting their prompts based on the descriptions in their respective papers. All baseline methods are evaluated using a fixed few-shot prompting setup to reduce prompt sensitivity.


The baselines we include are: \textit{Standard Prompting}, where the LLM generates answers without retrieval; \textit{Standard RAG } \cite{lewis2020retrieval}, which concatenates a fixed set of retrieved passages with the question; \textit{CoT} \cite{wei2022chain}, which guides the model to produce explicit reasoning steps; \textit{ReAct} \cite{yao2023react}, which interleaves reasoning steps with external actions; \textit{Self-Ask} \cite{press2022measuring}, which decomposes complex queries into sub-questions; \textit{IR-CoT} \cite{trivedi2022interleaving}, which interleaves retrieval with CoT reasoning to incrementally gather evidence during multi-hop inference; \textit{FLARE}  \cite{jiang2023active}, which adaptively triggers retrieval based on the model's uncertainty during generation; and \textit{Reflexion} \cite{shinn2023reflexion}, which enables iterative self-critique and refinement of answers. In addition to the baselines, we include \textit{SIM-RAG} \cite{yang2025knowing} not as a baseline but as a comparative framework, since it represents a RAG system that explicitly integrates metacognitive principles. SIM-RAG was introduced at SIGIR 2025, and we use the official implementation released by the authors, with minor adaptations to ensure a fair comparison with our MetaRAG setup.

\subsection{Experimental Setup}

We follow the experimental protocol described in the MetaRAG paper, with some adaptations to ensure clarity and reproducibility. This section details the LLMs used, retrieval and metacognitive configurations, reranking strategies, and the comparison with SIM-RAG.

\subsubsection{Language Models} We use \texttt{gpt-3.5-turbo-16k}\footnote{\url{https://platform.openai.com/docs/models}} (GPT-3.5) as the LLM, accessed through the OpenAI API\footnote{\url{https://platform.openai.com/docs/overview}}. The same LLM serves as the Reasoner model in both the CS and the MS. This corresponds to the closed-source model adopted in the original MetaRAG paper. Additionally, to assess generalization to open-source LLMs, we use Llama-3.3:70B\footnote{\url{https://ollama.com/library/llama3.3:70b}} (Llama3.3) as the Reasoner model across both spaces. All Llama3.3 experiments are run on a server equipped with four NVIDIA A100 SXM4 GPUs (40 GB VRAM each) and 514 GB of system memory, enabling efficient inference in a quantized setup. Both LLMs are configured with a temperature of 0.

\subsubsection{Retrieval Setup}
The original MetaRAG paper describes a hybrid retrieval setup that combines BM25 \cite{robertson2009probabilistic} as a sparse retriever with E5 \cite{wang2022text} as a dense retriever. However, neither the paper nor the released codebase provides sufficient implementation details to fully reproduce how these two retrieval signals are combined. To remain faithful to the intended hybrid design while ensuring reproducibility, we implement sparse--dense fusion from scratch using Reciprocal Rank Fusion (RRF) \cite{cormack2009reciprocal}. Sparse retrieval is performed using BM25 implemented in Lucene via Pyserini\footnote{\url{https://github.com/castorini/pyserini}}, while dense retrieval relies on E5 embeddings indexed with FAISS\footnote{\url{https://github.com/facebookresearch/faiss}}. Both retrievers are built over the same Wikipedia dump \cite{karpukhin2020dense}, which serves as the shared document collection for indexing. For each question, we independently retrieve the top 100 documents from the sparse and dense retrievers and merge the resulting ranked lists using RRF. After fusion, we retain the top five passages, which are passed to the answer generation and downstream reasoning components.


While the original SIM-RAG \cite{yang2025knowing} implementation employs a custom retriever and corpus built with ElasticSearch\footnote{\url{https://github.com/elastic/elasticsearch}}, we replace this component with our Pyserini--FAISS hybrid retrieval pipeline. This modification guarantees that both MetaRAG and SIM-RAG operate under identical retrieval conditions.

\subsubsection{Metacognitive Space Configuration}

In the MS, we reproduce the configuration described in the MetaRAG paper. A fine-tuned T5-large\footnote{\url{https://huggingface.co/gaussalgo/T5-LM-Large_Canard-HotpotQA-rephrase}} model is used as the expert for the monitoring stage. The judgment threshold is set to 0.4, which triggers the evaluation phase and allows a maximum of five iterations per question. For the evaluation and planning stages from the MetaRAG framework, we employ the same NLI T5-XXL\footnote{\url{https://huggingface.co/google/t5_xxl_true_nli_mixture}} model used in the original study. This configuration enables MetaRAG to assess evidence sufficiency, detect reasoning errors, and decide whether additional retrieval or reasoning refinement is required.

\subsubsection{Reranking Strategies}

To analyze the impact of reranking on MetaRAG, we evaluate two reranking configurations: \textit{PointWise} and \textit{ListWise} rerankers. Reranking is applied after each retrieval round, and the top five reranked documents are concatenated and passed to the LLM for answer generation. In the PointWise setting, each document–question pair is scored independently by the reranker, and the top-\textit{k} passages are selected based on their individual relevance scores. We evaluate three widely used PointWise rerankers: \textit{MiniLM}\footnote{\url{https://huggingface.co/cross-encoder/ms-marco-MiniLM-L12-v2}}, \textit{BGE}\footnote{\url{https://huggingface.co/BAAI/bge-reranker-v2-m3}} \cite{chen2024bge}, and \textit{ModernBERT}\footnote{\url{https://huggingface.co/Alibaba-NLP/gte-reranker-modernbert-base}} \cite{li2023towards,zhang2024mgte}. In contrast, the ListWise configuration jointly evaluates the full set of retrieved passages, allowing the model to consider inter-document relationships during ranking. For this setting, we employ \textit{RankGPT} \cite{sun2023chatgpt}, \textit{Zephyr} \cite{pradeep2023rankzephyr}, and \textit{Vicuna} \cite{sharifymoghaddam2025rankllm}, implemented through the RankLLM\footnote{\url{https://github.com/castorini/rank\_llm}} library \cite{sharifymoghaddam2025rankllm}, which provides ListWise reranking interfaces for LLMs. In our setup, RankGPT uses \texttt{gpt4o\_mini}\footnote{\url{https://platform.openai.com/docs/models/gpt-4o-mini}} as the underlying LLM for ListWise reranking. We understand that ListWise reranking introduces additional computational cost compared to PointWise or retrieval-only setups, particularly for RankGPT, which relies on a proprietary LLM. However, since our goal is to analyze the impact of reranking on MetaRAG rather than optimize for efficiency, we do not prioritize computational cost in this setup.

\subsubsection{SIM-RAG Critic Model}

For the SIM-RAG critic, we use \textit{SIM-RAG-Llama3-2B}\footnote{\url{https://huggingface.co/dyang39/SIM-RAG-Llama3-2B}}, a 2B-parameter Llama 3\footnote{\url{https://huggingface.co/andrijdavid/Llama3-2B-Base}} model adapted for general-purpose inference. It is fine tuned on several multi-hop QA benchmarks, including TriviaQA \cite{joshi2017triviaqa}, HotpotQA \cite{yang2018hotpotqa}, 2WikiMultiHopQA \cite{ho2020constructing}, PopQA \cite{rabinovich2023predicting}, and Musique \cite{trivedi2022interleaving}. 

\section{Results and Discussion}


This section presents our reproduced results for MetaRAG and discusses how they differ from the original study. We then examine diagnostic signals, threshold sensitivity, the impact of rerankers, and a comparison between MetaRAG and SIM-RAG in terms of accuracy and efficiency.

\subsection{Reproducibility of MetaRAG} 

To answer \textbf{RQ1}, we first investigate to what extent the results of MetaRAG can be reproduced under a setup that closely follows the configuration reported in the original paper. Our reproduced results show that, although absolute performance is generally lower than that reported in the original study, the relative trends are preserved. In particular, MetaRAG consistently outperforms all baseline methods across both HotpotQA and 2WikiMultiHopQA, including self-critique approaches such as Reflexion, as it is depicted in Table \ref{tab:metarag_repro_result}. This indicates that the core monitor--evaluate--plan mechanism proposed in MetaRAG remains effective under reproduced conditions. However, we observe some differences in absolute performance, particularly on 2WikiMultiHopQA, where the gap between the original and reproduced results is larger than on HotpotQA. Additionally, several baselines also achieve higher scores than originally reported, which may reflect changes in the underlying GPT-3.5 model over time due to deprecations and improvements\footnote{\url{https://platform.openai.com/docs/deprecations\#2023-11-06-chat-model-updates}}, introducing additional variability that hinders reproducibility with proprietary LLMs. We observe that Self-Ask performs worse than Standard RAG in our reproduced GPT-3.5 experiments, which differs from the results reported in prior work. Rather than indicating an implementation error, the discrepancy may stem from the fact that the Self-Ask baseline prompt used in the original MetaRAG experiments is not publicly released. Consequently, we attempt to approximate the prompt based on the description provided in the Self-Ask paper \cite{press2022measuring}. Since LLMs are known to be highly sensitive to even small changes in prompt wording and formatting ~\cite{schulhoff2024prompt,laskar2024systematic,razavi2025benchmarking}, such differences can influence how sub-questions are generated and how retrieval is triggered, ultimately affecting final answer quality.


\begin{table*}[t]
\centering
\setlength{\tabcolsep}{4pt}
\renewcommand{\arraystretch}{1.15}

\begin{tabular}{l l ccc cccc cccc}
\toprule
\multirow{2}{*}{\textbf{Data}} & \multirow{2}{*}{\textbf{Method}}
& \multirow{2}{*}{\textbf{Retr.}} & \multirow{2}{*}{\textbf{Multi.}} & \multirow{2}{*}{\textbf{Critic}}
& \multicolumn{4}{c}{\textbf{Original}}
& \multicolumn{4}{c}{\textbf{Reproduced}} \\
\cmidrule(lr){6-9} \cmidrule(lr){10-13}
& & & & & \textbf{EM} & \textbf{F1} & \textbf{Prec.} & \textbf{Rec.}
              & \textbf{EM} & \textbf{F1} & \textbf{Prec.} & \textbf{Rec.} \\
\midrule

\multirow{9}{*}{\rotatebox[origin=c]{90}{\textbf{HotpotQA}}}
& Standard Prompting & $\xmark$ & $\xmark$ & $\xmark$
& 20.0 & 25.8 & 26.4 & 28.9
& 26.4* & 36.6* & 39.4* & 35.2* \\
& Chain of Thought & $\xmark$ & $\xmark$ & $\xmark$
& 22.4 & 34.2 & 33.9 & 46.0
& 26.6* & 36.7* & 39.3* & 38.3* \\
\cmidrule(lr){2-13}
& Standard RAG & $\xmark$ & $\xmark$ & $\xmark$
& 24.6 & 33.0 & 34.1 & 34.5
& 30.8* & 42.0* & 45.1* & 42.2* \\
& ReAct & $\xmark$ & $\xmark$ & $\xmark$
& 24.8 & 41.7 & 42.6 & 44.7
& 31.8* & 42.4* & 46.8* & 41.3* \\
& Self-Ask & $\xmark$ & $\xmark$ & $\xmark$
& 28.2 & 43.1 & 43.4 & 44.8
& 28.6* & 39.2* & 41.2* & 43.1* \\
& FLARE & $\xmark$ & $\xmark$ & $\xmark$
& 29.2 & 42.4 & 42.8 & 43.0
& 33.6* & 44.7* & 48.4* & 44.0* \\
& IRCoT & $\xmark$ & $\xmark$ & $\xmark$
& 31.4 & 40.3 & 41.6 & 41.2
& 32.8* & 47.4* & 51.7* & 43.5* \\
\cmidrule(lr){2-13}
& Reflexion & \cmark & \cmark & \cmark
& 30.0 & 43.4 & 43.2 & 44.3
& 30.8* & 41.3* & 43.5* & 41.7* \\
& MetaRAG & \cmark & \cmark & \cmark
& \textbf{37.8} & \textbf{49.9} & \textbf{52.1} & \textbf{50.9}
& \textbf{34.2} & \textbf{48.2} & \textbf{49.5} & \textbf{48.6} \\

\midrule

\multirow{9}{*}{\rotatebox[origin=c]{90}{\textbf{2WikiMultiHopQA}}}
& Standard Prompting & $\xmark$ & $\xmark$ & $\xmark$
& 21.6 & 25.7 & 24.5 & 31.8
& 25.6* & 29.4* & 29.9* & 29.4* \\
& Chain of Thought & $\xmark$ & $\xmark$ & $\xmark$
& 27.6 & 37.4 & 35.8 & 44.3
& 25.8* & 31.2* & 31.7* & 32.0* \\
\cmidrule(lr){2-13}
& Standard RAG & $\xmark$ & $\xmark$ & $\xmark$
& 18.8 & 25.3 & 25.6 & 26.2
& 28.9* & 33.5* & 34.5* & 33.6* \\
& ReAct & $\xmark$ & $\xmark$ & $\xmark$
& 21.0 & 28.0 & 27.6 & 30.0
& 31.0* & 38.8* & 38.6 & 39.3*  \\
& Self-Ask & $\xmark$ & $\xmark$ & $\xmark$
& 28.6 & 37.5 & 36.5 & 42.8
& 27.2* & 34.6* & 35.8* & 34.9*  \\
& FLARE & $\xmark$ & $\xmark$ & $\xmark$
& 28.2 & 39.8 & 40.0 & 40.8
& 29.8* & \textbf{40.4} & 41.2* & \textbf{40.8} \\
& IRCoT & $\xmark$ & $\xmark$ & $\xmark$
& 30.8 & 42.6 & 42.3 & 40.9
& 30.6* & 39.9 & \textbf{43.1} & 38.2* \\
\cmidrule(lr){2-13}
& Reflexion & \cmark & \cmark & \cmark
& 31.8 & 41.7 & 40.6 & 44.2
& 31.2* & 36.8* & 35.9* & 37.8* \\
& MetaRAG & \cmark & \cmark & \cmark
& \textbf{42.8} & \textbf{50.8} & \textbf{50.7} & \textbf{52.2}
& \textbf{33.4} & 38.9 & 37.1 & 39.4 \\

\bottomrule
\end{tabular}

\caption{Evaluation results with retrieval (Retr.), multi-round retrieval (Multi.), and critic (Critic).
Baselines include Standard Prompting, Standard RAG, CoT, ReAct, Self-Ask, FLARE, IRCoT, and Reflexion. Original (reported in MetaRAG) vs. Reproduced (ours) results.
Best scores are in bold. An asterisk (*) denotes a statistically significant difference from MetaRAG ($p < 0.05$).}
\label{tab:metarag_repro_result}
\end{table*}

\subsection{Diagnostic Analysis}

The difference between the original and reproduced results, particularly on 2WikiMultiHopQA, motivates a deeper diagnostic analysis. We examine MetaRAG’s evaluation signals on HotpotQA and 2WikiMultiHopQA, including: (i) an LLM-based assessment of whether a question can be answered using internal knowledge alone, (ii)~an auxiliary LLM-based judgment of whether the retrieved documents are sufficient, and (iii) the NLI critic used to verify whether the retrieved documents entail the generated answer. Table~\ref{tab:metarag_diagnostics} highlights clear differences between the two datasets. While HotpotQA contains a noticeable portion of questions answerable using internal knowledge, 2WikiMultiHopQA is almost entirely retrieval-dependent. Moreover, when retrieval is required, the LLM more often judges the retrieved documents sufficient on HotpotQA than on 2WikiMultiHopQA, indicating that the initial top-\textit{k} retrieval documents more frequently fails to provide clearly sufficient evidence on 2WikiMultiHopQA.

In contrast, the NLI critic validates a larger fraction of answers than the LLM’s sufficiency judgment on both datasets. This suggests that relevant evidence is frequently present in the retrieved context but not easily recognized as sufficient prior to multi-hop reasoning by LLMs. The effect is particularly pronounced on 2WikiMultiHopQA, where many questions are not clearly solvable with the current evidence, so the system hesitates between stopping and continuing the reasoning process.

\begin{table*}[t]
\centering
\setlength{\tabcolsep}{8pt}
\renewcommand{\arraystretch}{1.15}

\begin{tabular}{lcc}
\toprule
\textbf{Evaluation Condition} & \textbf{HotpotQA} & \textbf{2WikiMultiHopQA} \\
\midrule

\multicolumn{3}{l}{\textit{All questions (LLM Judge)}} \\

Answered using internal knowledge & 20.4\% & 1.2\% \\
Requires external retrieval & 79.6\% & 98.8\% \\

\midrule
\multicolumn{3}{l}{\textit{Conditional on cases that require external retrieval (LLM Judge)}} \\

Answered using retrieved documents & 63.3\% & 33.6\% \\
Rejected despite retrieved documents & 36.7\% & 66.4\% \\

\midrule
\multicolumn{3}{l}{\textit{Conditional on cases that require external retrieval (NLI critic)}} \\

Validated by critic model & 86.7\% & 72.9\% \\
Rejected by critic  model & 13.3\% & 27.1\% \\

\bottomrule
\end{tabular}

\caption{Diagnostic breakdown of MetaRAG evaluation signals on 500-question subsets. The first block is computed over all 500 questions. The second and third blocks focus on questions the LLM marked as requiring retrieval: the second reports the LLM’s own assessment of retrieval sufficiency, and the third shows whether the NLI critic determines that the retrieved documents entail the generated answer (entailment = validate, otherwise = reject).}
\label{tab:metarag_diagnostics}
\end{table*}

\subsection{Sensitivity to the Metacognitive Threshold}



Our diagnostic analysis in Table \ref{tab:metarag_diagnostics} shows that nearly all questions in both datasets require external retrieval, with a notably larger gap on 2WikiMultiHopQA. Since MetaRAG relies on a confidence threshold to decide when to transition into the MS, this parameter controls how often additional retrieval and refinement are triggered. Given the high proportion of retrieval-dependent cases, this control mechanism may substantially affect performance. To examine its impact, we vary the threshold on the same 500-question subset of 2WikiMultiHopQA using GPT-3.5, while keeping all other components fixed. The results are reported in Table \ref{tab:threshold_sensitivity_2wiki}. We find that MetaRAG is highly sensitive to this parameter, as lower thresholds reduce transitions into the MS and additional retrieval calls, while improving answer quality. In the original MetaRAG paper, the threshold is set to 0.4, which yields the best results under their setup. In contrast, our reproduction, answer quality (e.g., EM/F1) peaks at a threshold of 0.2, whereas further lowering the threshold to 0.1 results in decreased performance, even though it reduces retrieval and generation calls. One possible explanation is that GPT-3.5 has evolved\footnote{\url{https://platform.openai.com/docs/deprecations\#2023-11-06-chat-model-updates}} since the original study, potentially strengthening its internal knowledge. As a result, the original threshold configuration may no longer be optimal, and repeatedly triggering metacognitive control may provide limited additional benefit.

Notably, this analysis does not contradict our earlier diagnostic findings in Table \ref{tab:metarag_diagnostics}. Although the LLM frequently judges the retrieved context as insufficient, the NLI critic validates a much larger portion of answers, indicating that the necessary evidence is often already present. The bottleneck therefore appears to lie in how evidence is connected across hops rather than in its availability. Therefore, increasing retrieval iterations alone does not necessarily improve reasoning, and may instead complicate the detection of relevant information.

For all subsequent experiments, we fix the confidence threshold at 0.4 in order to isolate the effects of reranking improvements over retrieval-only set up.

\begin{table}[t]
\centering
\setlength{\tabcolsep}{6pt}
\renewcommand{\arraystretch}{1.15}
\begin{tabular}{l cc ccc}
\toprule
\textbf{Threshold} & \textbf{EM} & \textbf{F1} &
\textbf{MS trans./q} & \textbf{Retr./q} & \textbf{Iters./q} \\
\midrule
0.4 & 33.4 & 38.9 & 3.2 & 2.8 & 4.2 \\
0.3 & 37.8 & 45.6 & 2.5 & 2.1 & 3.5 \\
0.2 & 38.1 & 46.2 & 1.5 & 1.2 & 2.5 \\
0.1 & 35.4 & 42.6 & 1.3 & 1.1 & 2.2 \\
\bottomrule
\end{tabular}
\caption{Sensitivity to the confidence threshold on 2WikiMultiHopQA (500 samples; GPT-3.5; max\_iter=5; top-$k$=5). The parameter max\_iter denotes the maximum number of MetaRAG reasoning iterations permitted per question, whereas top-$k$ defines the number of documents retrieved at each retriever call. The columns MS trans./q, Retr./q, and Iters./q denote the average metacognitive transitions, retrieval calls, and reasoning iterations per question, respectively. Results are reported as Exact Match (EM) and F1.}

\label{tab:threshold_sensitivity_2wiki}
\end{table}

\subsection{Enhancing MetaRAG with Rerankers}

To answer \textbf{RQ2}, we examine whether reranker models improve MetaRAG's performance by promoting more relevant documents within the top-\textit{k} context used for reasoning and critique. As shown in Table~\ref{tab:rerankers_metarag}, reranking generally improves performance on both collections; however, its effectiveness strongly depends on the specific reranker configuration. Overall, BGE delivers the strongest gains among PointWise methods, while RankGPT performs best among ListWise approaches. Moreover, reranking yields larger improvements with Llama3.3 than with GPT-3.5, suggesting that reranking can improve document ordering, although its effectiveness may rely on the LLM’s ability to connect reasoning hops across the reordered evidence and generate an accurate answer from it.

Reranking is not uniformly beneficial and, in certain cases, performs worse than the retrieval-only setup. For example, MiniLM reduces performance on HotpotQA when paired with Llama3.3 and similarly underperforms on both collections when used with GPT-3.5. This pattern is consistent with limitations of PointWise reranking in multi-hop settings. PointWise rerankers evaluate each document independently with respect to the question, which may be suboptimal when answering requires combining evidence across multiple passages. MiniLM is trained on the MS MARCO passage collection \cite{nguyen2016ms}, which primarily contains single-hop question–passage pairs. As a result, it may prioritize documents that are individually similar to the question rather than those that collectively support multi-hop reasoning across multiple sources. In contrast, BGE is trained on a larger and more diverse mixture of retrieval and QA-oriented datasets \cite{chen2024bge}, which enables it to better capture deeper semantic relevance. Similarly, RankGPT’s strong performance among ListWise approaches may stem from its ability to consider all candidate documents jointly when producing a ranking. Unlike PointWise methods, ListWise rerankers model interactions between passages, which is particularly advantageous in multi-hop question answering where evidence is combined across several documents.

\begin{table*}[t]
\centering
\setlength{\tabcolsep}{4pt}
\renewcommand{\arraystretch}{1.15}

\begin{tabular}{l l l cccc cccc}
\toprule
\multirow{2}{*}{\textbf{LLM}} &
\multirow{2}{*}{\textbf{Setting}} &
\multirow{2}{*}{\textbf{Reranker}} &
\multicolumn{4}{c}{\textbf{HotpotQA}} &
\multicolumn{4}{c}{\textbf{2WikiMultiHopQA}} \\
\cmidrule(lr){4-7} \cmidrule(lr){8-11}
& & & \textbf{EM} & \textbf{F1} & \textbf{Prec.} & \textbf{Rec.}
      & \textbf{EM} & \textbf{F1} & \textbf{Prec.} & \textbf{Rec.} \\
\midrule

\multirow{7}{*}{\rotatebox[origin=c]{90}{\textbf{GPT-3.5}}}
& Ret & -- 
& 34.2 & 48.2 & 49.5 & 48.6
& 33.4 & 38.9 & 37.1 & 39.4 \\
\cmidrule(lr){2-11}

& \multirow{3}{*}{Ret+PointWise} & MiniLM
& 34.1 & 48.4 & 50.1 & 47.3
& 31.9 & 36.8 & 37.5* & 36.3 \\
&  & BGE
& 38.8* & 51.3* & 53.3* & 51.5*
& 36.3* & 42.1* & 43.1* & 42.4* \\
&  & ModernBERT
& 35.7* & 48.7* & 52.4* & 49.6*
& 30.9 & 35.1 & 36.7 & 36.4 \\
\cmidrule(lr){2-11}

& \multirow{3}{*}{Ret+ListWise} & RankGPT
& \textbf{40.1*} & \textbf{53.5*} & \textbf{55.5*} & \textbf{54.2*}
& \textbf{38.9*} & \textbf{44.3*} & \textbf{45.4*} & \textbf{43.6*} \\
&  & Zephyr
& 36.3* & 46.0 & 47.2 & 49.6*
& 32.0 & 34.0 & 32.9 & 33.8 \\
&  & Vicuna
& 35.8* & 48.6* & 51.4* & 46.5
& 33.9* & 40.8* & 40.1* & 41.9* \\

\midrule

\multirow{7}{*}{\rotatebox[origin=c]{90}{\textbf{Llama3.3}}}
& Ret & --
& 41.5 & 54.1 & 57.9 & 55.6
& 29.0 & 33.3 & 31.3 & 32.9 \\
\cmidrule(lr){2-11}

& \multirow{3}{*}{Ret+PointWise} & MiniLM
& 40.4 & 51.3 & 56.1 & 52.1
& 33.7* & 40.1* & 41.5* & 42.1* \\
&  & BGE
& 43.6* & 56.1* & 61.3* & 56.1
& 35.5* & 41.3* & 42.1* & 42.9* \\
&  & ModernBERT
& 41.8 & 54.7* & 59.4* & 53.9
& 33.2* & 36.9* & 37.9* & 36.8* \\
\cmidrule(lr){2-11}

& \multirow{3}{*}{Ret+ListWise} & RankGPT
& \textbf{46.7*} & \textbf{59.3*} & \textbf{63.1*} & \textbf{57.1*}
& \textbf{38.9*} & \textbf{45.5*} & \textbf{48.6*} & \textbf{46.7*} \\
&  & Zephyr
& 44.4* & 58.8* & 62.1* & 56.3
& 33.7* & 37.9* & 35.4* & 35.2* \\
&  & Vicuna
& 41.2 & 53.5 & 58.9 & 51.4
& 36.2* & 40.8* & 39.2* & 38.1* \\

\bottomrule
\end{tabular}

\caption{Comparison of MetaRAG performance under retrieval-only (Ret), PointWise, and ListWise reranking settings across different LLMs and datasets.
Results are reported as EM, F1, Precision (Prec.), and Recall (Rec.).
Best scores are in bold. An asterisk (*) denotes a statistically significant improvement ($p<0.05$) over the corresponding retrieval-only baseline within the same LLM.}
\label{tab:rerankers_metarag}
\end{table*}

\subsection{Comparison with SIM-RAG}

\subsubsection{Performance Comparison}

In \textbf{RQ3}, we compare MetaRAG against SIM-RAG to study its performance relative to other metacognitive frameworks. We evaluate both under identical retrieval and reranking conditions, as shown in Table~\ref{tab:simrag_comparison}. Our findings show that MetaRAG performs significantly better than SIM-RAG when supported by reranker models. For this comparison, we select the best PointWise and ListWise settings from the previous experiment, namely BGE and RankGPT, respectively. The results reveal a clear pattern across both GPT3.5 and Llama3.3: while SIM-RAG performs reasonably well in retrieval-only settings, its performance drops once rerankers are introduced, especially with Llama3.3, though the trend is also present for GPT3.5. This may be because SIM-RAG's fine-tuned critic is tailored to a fixed retrieval setup, and rerankers modify the input distribution, disrupting the conditions under which it operates effectively. In contrast, MetaRAG conducts self-critique through prompting rather than fine-tuning, making it more flexible and less sensitive to such distributional changes, and thus better able to adapt to varying retrieval conditions. Paired t-tests are conducted between MetaRAG and SIM-RAG. Several settings show statistically significant improvements in favor of MetaRAG ($p < 0.05$), as indicated by the asterisks in Table~\ref{tab:simrag_comparison}, while other configurations are favor SIM-RAG.

\begin{table*}[t]
\centering
\setlength{\tabcolsep}{4pt}
\renewcommand{\arraystretch}{1.15}

\begin{tabular}{l l c l cccc cccc}
\toprule
\textbf{Method} & \textbf{LLM} & \textbf{Setting} & \textbf{Reranker}
& \multicolumn{4}{c}{\textbf{HotpotQA}}
& \multicolumn{4}{c}{\textbf{2WikiMultiHopQA}} \\
\cmidrule(lr){5-8} \cmidrule(lr){9-12}
& & & & \textbf{EM} & \textbf{F1} & \textbf{Prec.} & \textbf{Rec.}
      & \textbf{EM} & \textbf{F1} & \textbf{Prec.} & \textbf{Rec.} \\
\midrule

\multirow{6}{*}{\textbf{MetaRAG}}
& \multirow{3}{*}{GPT-3.5}
& Ret & --
& 34.2 & 48.2 & 49.5 & 48.6 
& 33.4* & 38.9* & 37.1* & 39.4* \\
&  & Ret+PointWise & BGE
& 38.8 & 51.3* & 53.3* & 51.5*
& \textbf{36.3} & \textbf{42.1} & \textbf{43.1} & \textbf{42.4} \\
&  & Ret+ListWise & RankGPT
& \textbf{40.1} & \textbf{53.5*} & \textbf{55.5*} & \textbf{54.2*}
& 35.9* & 41.3* & 42.4* & 40.6* \\
\cmidrule(lr){2-12}

& \multirow{3}{*}{Llama3.3}
& Ret & --
& 41.5 & 54.1* & 57.9* & 55.6*
& 29.0 & 33.3 & 31.3 & 32.9 \\
&  & Ret+PointWise & BGE
& 43.6* & 56.1* & 61.3* & 56.1*
& 35.5* & 41.3* & 42.1* & 42.9* \\
&  & Ret+ListWise & RankGPT
& \textbf{46.7*} & \textbf{59.3*} & \textbf{63.1*} & \textbf{57.1*}
& \textbf{38.9*} & \textbf{45.5*} & \textbf{48.6*} & \textbf{46.7*} \\

\midrule

\multirow{6}{*}{\textbf{SIM-RAG}}
& \multirow{3}{*}{GPT-3.5}
& Ret & --
& \textbf{42.2} & \textbf{53.1} & \textbf{55.4} & \textbf{54.9} 
& 30.8 & 34.4 & 35.3 & 33.7 \\
&  & Ret+PointWise & BGE
& 41.6 & 44.3 & 46.3 & 45.5
& \textbf{36.3} & \textbf{45.6} & \textbf{49.2} & \textbf{47.3} \\
&  & Ret+ListWise & RankGPT
& 41.2 & 47.5 & 49.5 & 50.1
& 23.5 & 25.3 & 29.4 & 27.4 \\
\cmidrule(lr){2-12}

& \multirow{3}{*}{Llama3.3}
& Ret & --
& \textbf{43.1} & \textbf{60.9} & \textbf{61.3} & \textbf{58.4}
& \textbf{38.3} & \textbf{43.3} & \textbf{45.6} & \textbf{43.2} \\
&  & Ret+PointWise & BGE
& 21.4 & 25.0 & 26.4 & 26.3
& 14.5 & 15.1 & 15.6 & 14.9 \\
&  & Ret+ListWise & RankGPT
& 18.7 & 22.3 & 23.8 & 22.5
& 12.8 & 16.5 & 19.3 & 18.9 \\

\bottomrule
\end{tabular}

\caption{Comparison of MetaRAG and SIM-RAG performance across retrieval-only (Ret), PointWise, and ListWise reranking settings.
The \textit{Reranker} column specifies the reranking model used.
Results are reported as Exact Match (EM), F1, Precision (Prec.), and Recall (Rec.).
Best results are in bold.
An asterisk (*) indicates MetaRAG significantly outperforms SIM-RAG under the same LLM and setting ($p<0.05$).}
\label{tab:simrag_comparison}
\end{table*}

\subsubsection{Efficiency and Computational Cost}


Beyond accuracy, we compare MetaRAG and SIM-RAG in terms of computational cost under the retrieval-only setup. As shown in Table ~\ref{tab:resources_hotpotqa}~(a) and Table~\ref{tab:resources_hotpotqa}~(b), MetaRAG incurs substantially higher overhead across both LLMs. It requires more model calls per question and exhibits significantly higher token usage and latency than SIM-RAG, resulting in a markedly higher overall cost in the GPT-3.5 setting. Similar trends are observed with Llama3.3, where MetaRAG’s iterative control mechanism leads to increased request counts and slower inference. Overall, the results highlight a clear efficiency gap between the two frameworks. However, while MetaRAG requires considerably greater computational resources to reach high accuracy, it remains training-free and broadly applicable across datasets, whereas SIM-RAG achieves faster and more resource-efficient inference but depends on dataset-specific critic fine-tuning, potentially constraining its cross-domain generalization despite its efficiency gains

\begin{table}[t]
\centering
\small

\textbf{(a) GPT-3.5}

\vspace{0.3em}

\begin{tabular}{l
                S[table-format=2.2]
                S[table-format=5.0]
                S[table-format=2.2]
                S[table-format=1.2]}
\toprule
\textbf{Method} &
{\textbf{Req./question}} &
{\textbf{Tok./req.}} &
{\textbf{Lat./req. (s)}} &
{\textbf{Cost (USD)}} \\
\midrule
RAG     & {1.0}  & {870.0}    & {0.5}  & {1.3} \\
MetaRAG & {14.85} & {30,064.5} & {47.85} & {37.66} \\
SIM-RAG  & {6.5}  & {14,118.4}    & {7.2}  & {21.3} \\
\bottomrule
\end{tabular}

\vspace{0.8em}


\textbf{(b) Llama3.3}

\vspace{0.3em}

\begin{tabular}{l
                S[table-format=2.2]
                S[table-format=5.0]
                S[table-format=2.2]
                S[table-format=1.2]}
\toprule
\textbf{Method} &
{\textbf{Req./question}} &
{\textbf{Tok./req.}} &
{\textbf{Lat./req. (s)}} &
{\textbf{Cost (USD)}} \\
\midrule
RAG     & {1.0}  & {871.7}    & {1.7}  & \multicolumn{1}{c}{--} \\
MetaRAG & {18.9} & {26,004.6}   & {121.4} & \multicolumn{1}{c}{--} \\
SIM-RAG  & {14.1}  & {31,107.5}    & {103.8}  & \multicolumn{1}{c}{--} \\
\bottomrule
\end{tabular}
\caption{Computational cost on HotpotQA (500 questions) under the retrieval-only setup. (a) GPT-3.5 and (b) Llama3.3 report the average number of model requests per question (Req./question),  average tokens per request (Tok./req.), average latency per request in seconds (Lat./req.), and total monetary cost (USD) for processing the full dataset.}
\label{tab:resources_hotpotqa}
\end{table}

\section{Reproducibility Challenges}

We identify several challenges that explain the discrepancies between our results and those reported in the original paper. 

First, the use of closed-source LLMs, such as those from the OpenAI family, hinders reproducibility because these models are continuously updated to provide better answers for users and, consequently, do not produce identical outputs over time \cite{laskar2024systematic}. For instance, the model GPT3.5 undergoes several updates after its release\footnote{\url{https://platform.openai.com/docs/deprecations\#2023-11-06-chat-model-updates}}, which makes exact numerical reproduction impossible even when following the same methodology as the original authors.

Second, the original paper describes a hybrid retrieval strategy combining sparse (BM25) and dense (E5) methods, but it does not specify how these signals are integrated, nor is this in any detail reflected in the released code. In our implementation, we explicitly construct both components: we use BM25 via Pyserini\footnote{https://github.com/castorini/pyserini} and build a dense retriever by encoding the Wikipedia dump with E5 embeddings and indexing them using FAISS. The two retrieval outputs are fused using RRF. Since these implementation details are absent from the original paper and code, differences in the hybrid retrieval setup may partly explain the lower absolute performance observed in our results.
 
Third, neither the prompts nor the baseline implementations are released. As a consequence, we re-implement both from scratch, which hinders reproducibility and makes it difficult to obtain the same scores for each baseline. Although our implementations closely follow the descriptions provided in the original paper, even small changes to prompts can significantly affect LLM outputs and evaluation metric results.

Fourth, the original paper reports results based on a subset of 500 QA pairs from the development split of each collection, but this subset is not shared in the released code. To reproduce the experiments, we randomly sample 500 QA pairs from the respective development sets of HotpotQA and 2WikiMultiHopQA collections. This difference in sample selection likely contributes to the score variations between our results and those reported in the original MetaRAG paper.

Fifth, the original paper lacks an explanation of how the index for fetching passages from Wikipedia is constructed. The authors state that they use ElasticSearch for indexing, whereas our approach relies on Pyserini, which provides a more transparent and easily reproducible interface to Lucene. 
Although both systems are built on the same underlying Lucene engine, differences in configuration and implementation details can influence which documents are retrieved and, consequently, affect the generated answers~\cite{kamphuis2020bm25}.

Sixth, the released code contains evidence of reranker usage, specifically the model \texttt{intfloat/simlm-msmarco-reranker}\footnote{\url{https://github.com/ignorejjj/MetaRAG/blob/bd85d13cb3500afc119e178500ea9dbaee4d99e5/config.py\#L18}}. However, this reranker is neither mentioned nor discussed in the original paper, and its role within the MetaRAG pipeline remains unclear. Since the paper refers only broadly to hybrid retrieval, it is uncertain how this reranker is applied and how it impacts the reported results. This ambiguity motivates our further exploration of different retrieval and reranking setups to better understand their influence on metacognitive RAG frameworks such as MetaRAG and SIM-RAG.

Finally, we attempted to contact the authors on three occasions via email to clarify these discrepancies and better understand the released code. Unfortunately, we did not receive a response, which further limited our ability to resolve the identified reproducibility issues.

\section{Conclusions and Future Work}


In this work, we conduct a reproducibility study of MetaRAG for multi-hop QA. Our results confirm its main qualitative claim: MetaRAG consistently outperforms standard prompting, Standard RAG, and strong reasoning baselines such as CoT, ReAct, Self-Ask, IRCoT, FLARE, and Reflexion on both HotpotQA and 2WikiMultiHopQA. However, we observe lower absolute scores than those reported in the original paper, particularly on 2WikiMultiHopQA, highlighting the sensitivity of RAG systems to evolving closed-source LLMs and underspecified implementation details. Our diagnostic analysis shows that retrieval is required for most questions, yet the LLM often judges the top-k evidence insufficient even when an NLI critic later validates the answer. This suggests that performance limitations are more related to evidence ordering and multi-hop reasoning than to missing information in the retrieved documents. We also find that MetaRAG is sensitive to its judgment threshold, where lower values reduce metacognitive transitions and extra retrieval while improving accuracy in our reproduced setting.

We further extend MetaRAG with reranker models and observe consistent gains in both PointWise and ListWise settings, especially when paired with stronger LLM backends. In contrast, SIM-RAG, while competitive in retrieval-only configurations, does not consistently benefit from reranking and in s settings degrades subseveraltantially when the retrieved evidence is reordered or altered. This suggests that SIM-RAG’s fine-tuned critic is more sensitive to retrieval modifications, whereas MetaRAG’s prompt-based metacognitive control is more adaptable across configurations. Overall, our findings indicate that MetaRAG’s effectiveness depends strongly on key design choices, including the judgment threshold, retrieval strategy, reranking method, and underlying LLM. 



As future work, we aim to extend metacognitive RAG to additional knowledge-intensive tasks, such as biomedical question answering and mathematical reasoning, where balancing retrieval and reasoning is particularly critical. We also plan to compare metacognitive RAG frameworks with alternative approaches that learn retrieval and search behaviors via reinforcement learning, such as Search-R1 \cite{jin2025search}, as well as with other agent-based RAG systems. Moreover, we will broaden our evaluation beyond final answer accuracy by assessing the quality of retrieved evidence and the correctness and coherence of intermediate reasoning steps \cite{freja2026evalqreason}. 






\bibliographystyle{ACM-Reference-Format}
\balance
\bibliography{sigir2026}

@inproceedings{zhou2024metacognitive,
  title={Metacognitive retrieval-augmented large language models},
  author={Zhou, Yujia and Liu, Zheng and Jin, Jiajie and Nie, Jian-Yun and Dou, Zhicheng},
  booktitle={Proc. ACM Web Conf. 2024},
  pages={1453--1463},
  year={2024}
}

@inproceedings{izacard2021leveraging,
  title={Leveraging Passage Retrieval with Generative Models for Open Domain Question Answering},
  author={Izacard, Gautier and Grave, Edouard},
  booktitle={EACL 2021-16th Conference of the European Chapter of the Association for Computational Linguistics},
  pages={874--880},
  year={2021},
  organization={Association for Computational Linguistics}
}

@article{lewis2020retrieval,
  title={Retrieval-augmented generation for knowledge-intensive nlp tasks},
  author={Lewis, Patrick and Perez, Ethan and Piktus, Aleksandra and Petroni, Fabio and Karpukhin, Vladimir and Goyal, Naman and K{\"u}ttler, Heinrich and Lewis, Mike and Yih, Wen-tau and Rockt{\"a}schel, Tim and others},
  journal={Advances in neural information processing systems},
  volume={33},
  pages={9459--9474},
  year={2020}
}

@article{yang2018hotpotqa,
  title={HotpotQA: A dataset for diverse, explainable multi-hop question answering},
  author={Yang, Zhilin and Qi, Peng and Zhang, Saizheng and Bengio, Yoshua and Cohen, William W and Salakhutdinov, Ruslan and Manning, Christopher D},
  journal={arXiv preprint arXiv:1809.09600},
  year={2018}
}

@inproceedings{ho2020constructing,
  title={Constructing A Multi-hop QA Dataset for Comprehensive Evaluation of Reasoning Steps},
  author={Ho, Xanh and Nguyen, Anh-Khoa Duong and Sugawara, Saku and Aizawa, Akiko},
  booktitle={Proceedings of the 28th International Conference on Computational Linguistics},
  pages={6609--6625},
  year={2020}
}

@article{tang2024multihop,
  title={Multihop-rag: Benchmarking retrieval-augmented generation for multi-hop queries},
  author={Tang, Yixuan and Yang, Yi},
  journal={arXiv preprint arXiv:2401.15391},
  year={2024}
}

@article{bahak2023evaluating,
  title={Evaluating chatgpt as a question answering system: A comprehensive analysis and comparison with existing models},
  author={Bahak, Hossein and Taheri, Farzaneh and Zojaji, Zahra and Kazemi, Arefeh},
  journal={arXiv preprint arXiv:2312.07592},
  year={2023}
}

@inproceedings{yang2025knowing,
  title={Knowing You Don't Know: Learning When to Continue Search in Multi-round RAG through Self-Practicing},
  author={Yang, Diji and Zeng, Linda and Rao, Jinmeng and Zhang, Yi},
  booktitle={Proceedings of the 48th International ACM SIGIR Conference on Research and Development in Information Retrieval},
  pages={1305--1315},
  year={2025}
}

@article{wei2022chain,
  title={Chain-of-thought prompting elicits reasoning in large language models},
  author={Wei, Jason and Wang, Xuezhi and Schuurmans, Dale and Bosma, Maarten and Xia, Fei and Chi, Ed and Le, Quoc V and Zhou, Denny and others},
  journal={Advances in neural information processing systems},
  volume={35},
  pages={24824--24837},
  year={2022}
}

@inproceedings{yao2023react,
  title={React: Synergizing reasoning and acting in language models},
  author={Yao, Shunyu and Zhao, Jeffrey and Yu, Dian and Du, Nan and Shafran, Izhak and Narasimhan, Karthik and Cao, Yuan},
  booktitle={International Conference on Learning Representations (ICLR)},
  year={2023}
}

@article{press2022measuring,
  title={Measuring and narrowing the compositionality gap in language models},
  author={Press, Ofir and Zhang, Muru and Min, Sewon and Schmidt, Ludwig and Smith, Noah A and Lewis, Mike},
  journal={arXiv preprint arXiv:2210.03350},
  year={2022}
}

@article{shinn2023reflexion,
  title={Reflexion: Language agents with verbal reinforcement learning, 2023},
  author={Shinn, Noah and Cassano, Federico and Labash, Beck and Gopinath, Ashwin and Narasimhan, Karthik and Yao, Shunyu},
  journal={URL https://arxiv. org/abs/2303.11366},
  volume={1},
  year={2023}
}

@article{asai2024self,
  title={Self-rag: Learning to retrieve, generate, and critique through self-reflection},
  author={Asai, Akari and Wu, Zeqiu and Wang, Yizhong and Sil, Avirup and Hajishirzi, Hannaneh},
  year={2024},
  publisher={ICLR}
}

@article{trivedi2022interleaving,
  title={Interleaving retrieval with chain-of-thought reasoning for knowledge-intensive multi-step questions},
  author={Trivedi, Harsh and Balasubramanian, Niranjan and Khot, Tushar and Sabharwal, Ashish},
  journal={arXiv preprint arXiv:2212.10509},
  year={2022}
}

@inproceedings{jiang2023active,
  title={Active retrieval augmented generation},
  author={Jiang, Zhengbao and Xu, Frank F and Gao, Luyu and Sun, Zhiqing and Liu, Qian and Dwivedi-Yu, Jane and Yang, Yiming and Callan, Jamie and Neubig, Graham},
  booktitle={Proceedings of the 2023 Conference on Empirical Methods in Natural Language Processing},
  pages={7969--7992},
  year={2023}
}

@inproceedings{he2021efficient,
  title={Efficient Nearest Neighbor Language Models},
  author={He, Junxian and Neubig, Graham and Berg-Kirkpatrick, Taylor},
  booktitle={Conference on Empirical Methods in Natural Language Processing},
  year={2021}
}

@article{zhong2022training,
  title={Training language models with memory augmentation},
  author={Zhong, Zexuan and Lei, Tao and Chen, Danqi},
  journal={arXiv preprint arXiv:2205.12674},
  year={2022}
}

@inproceedings{khandelwalgeneralization,
  title={Generalization through Memorization: Nearest Neighbor Language Models},
  author={Khandelwal, Urvashi and Levy, Omer and Jurafsky, Dan and Zettlemoyer, Luke and Lewis, Mike},
  booktitle={International Conference on Learning Representations}
}

@inproceedings{trivedi2023interleaving,
  title={Interleaving Retrieval with Chain-of-Thought Reasoning for Knowledge-Intensive Multi-Step Questions},
  author={Trivedi, Harsh and Balasubramanian, Niranjan and Khot, Tushar and Sabharwal, Ashish},
  booktitle={Proceedings of the 61st Annual Meeting of the Association for Computational Linguistics (Volume 1: Long Papers)},
  pages={10014--10037},
  year={2023}
}

@inproceedings{khotdecomposed,
  title={Decomposed Prompting: A Modular Approach for Solving Complex Tasks},
  author={Khot, Tushar and Trivedi, Harsh and Finlayson, Matthew and Fu, Yao and Richardson, Kyle and Clark, Peter and Sabharwal, Ashish},
  booktitle={The Eleventh International Conference on Learning Representations}
}

@inproceedings{wang2022selfconsistency,
  title={Self-consistency improves chain of thought reasoning in language models},
  author={Wang, Xuezhi and Wei, Jason and Schuurmans, Dale and others},
  booktitle={International Conference on Learning Representations (ICLR)},
  year={2022}
}

@article{yu2025unleashing,
  title={Unleashing the Power of Context Repetition for Robust Reasoning in LLMs},
  author={Yu, Tianyu and others},
  journal={arXiv preprint arXiv:2503.06789},
  year={2025}
}

@inproceedings{kamphuis2020bm25,
  title={Which BM25 do you mean? A large-scale reproducibility study of scoring variants},
  author={Kamphuis, Chris and De Vries, Arjen P and Boytsov, Leonid and Lin, Jimmy},
  booktitle={European Conference on Information Retrieval},
  pages={28--34},
  year={2020},
  organization={Springer}
}

@article{lai2011metacognition,
  title={Metacognition: A literature review},
  author={Lai, Emily R},
  year={2011},
  publisher={Pearson Research Report. ER Lai--Pearson Education, 2011 [Electronic resource~…}
}

@article{sun2023chatgpt,
  title={Is ChatGPT good at search? investigating large language models as re-ranking agents},
  author={Sun, Weiwei and Yan, Lingyong and Ma, Xinyu and Wang, Shuaiqiang and Ren, Pengjie and Chen, Zhumin and Yin, Dawei and Ren, Zhaochun},
  journal={arXiv preprint arXiv:2304.09542},
  year={2023}
}

@article{jeong2024adaptive,
  title={Adaptive-rag: Learning to adapt retrieval-augmented large language models through question complexity},
  author={Jeong, Soyeong and Baek, Jinheon and Cho, Sukmin and Hwang, Sung Ju and Park, Jong C},
  journal={arXiv preprint arXiv:2403.14403},
  year={2024}
}

@article{li2024smart,
  title={SMART-RAG: Selection using Determinantal Matrices for Augmented Retrieval},
  author={Li, Jiatao and Hu, Xinyu and Wan, Xiaojun},
  journal={arXiv preprint arXiv:2409.13992},
  year={2024}
}

@article{pradeep2023rankzephyr,
  title={RankZephyr: Effective and Robust Zero-Shot Listwise Reranking is a Breeze!},
  author={Pradeep, Ronak and Sharifymoghaddam, Sahel and Lin, Jimmy},
  journal={arXiv preprint arXiv:2312.02724},
  year={2023}
}

@inproceedings{sharifymoghaddam2025rankllm,
  title={Rankllm: A python package for reranking with llms},
  author={Sharifymoghaddam, Sahel and Pradeep, Ronak and Slavescu, Andre and Nguyen, Ryan and Xu, Andrew and Chen, Zijian and Zhang, Yilin and Chen, Yidi and Xian, Jasper and Lin, Jimmy},
  booktitle={Proceedings of the 48th International ACM SIGIR Conference on Research and Development in Information Retrieval},
  pages={3681--3690},
  year={2025}
}

@inproceedings{karpukhin2020dense,
  title={Dense Passage Retrieval for Open-Domain Question Answering.},
  author={Karpukhin, Vladimir and Oguz, Barlas and Min, Sewon and Lewis, Patrick SH and Wu, Ledell and Edunov, Sergey and Chen, Danqi and Yih, Wen-tau},
  booktitle={EMNLP (1)},
  pages={6769--6781},
  year={2020}
}

@article{jin2025search,
  title={Search-r1: Training llms to reason and leverage search engines with reinforcement learning},
  author={Jin, Bowen and Zeng, Hansi and Yue, Zhenrui and Yoon, Jinsung and Arik, Sercan and Wang, Dong and Zamani, Hamed and Han, Jiawei},
  journal={arXiv preprint arXiv:2503.09516},
  year={2025}
}

@article{schraw1995metacognitive,
  title={Metacognitive theories},
  author={Schraw, Gregory and Moshman, David},
  journal={Educational psychology review},
  volume={7},
  number={4},
  pages={351--371},
  year={1995},
  publisher={Springer}
}

@article{li2024hindsight,
  title={When hindsight is not 20/20: Testing limits on reflective thinking in large language models},
  author={Li, Yanhong and Yang, Chenghao and Ettinger, Allyson},
  journal={arXiv preprint arXiv:2404.09129},
  year={2024}
}

@article{gotoh2016development,
  title={Development of Critical Thinking with Metacognitive Regulation.},
  author={Gotoh, Yasushi},
  journal={International association for development of the information society},
  year={2016},
  publisher={ERIC}
}

@misc{nelson1990metamemory,
  title={Metamemory: A theoretical framework and some new findings. The Psychology of Learning and Motivation. Vol. 26},
  author={Nelson, TO and Narens, L},
  year={1990},
  publisher={New York, NY: Academic Press}
}

@article{mavi2024multi,
  title={Multi-hop question answering},
  author={Mavi, Vaibhav and Jangra, Anubhav and Jatowt, Adam and others},
  journal={Foundations and Trends{\textregistered} in Information Retrieval},
  volume={17},
  number={5},
  pages={457--586},
  year={2024},
  publisher={Now Publishers, Inc.}
}

@misc{chen2024bge,
      title={BGE M3-Embedding: Multi-Lingual, Multi-Functionality, Multi-Granularity Text Embeddings Through Self-Knowledge Distillation}, 
      author={Jianlv Chen and Shitao Xiao and Peitian Zhang and Kun Luo and Defu Lian and Zheng Liu},
      year={2024},
      eprint={2402.03216},
      archivePrefix={arXiv},
      primaryClass={cs.CL}
}

@inproceedings{zhang2024mgte,
  title={mGTE: Generalized Long-Context Text Representation and Reranking Models for Multilingual Text Retrieval},
  author={Zhang, Xin and Zhang, Yanzhao and Long, Dingkun and Xie, Wen and Dai, Ziqi and Tang, Jialong and Lin, Huan and Yang, Baosong and Xie, Pengjun and Huang, Fei and others},
  booktitle={Proceedings of the 2024 Conference on Empirical Methods in Natural Language Processing: Industry Track},
  pages={1393--1412},
  year={2024}
}

@article{li2023towards,
  title={Towards general text embeddings with multi-stage contrastive learning},
  author={Li, Zehan and Zhang, Xin and Zhang, Yanzhao and Long, Dingkun and Xie, Pengjun and Zhang, Meishan},
  journal={arXiv preprint arXiv:2308.03281},
  year={2023}
}

@inproceedings{cormack2009reciprocal,
  title={Reciprocal rank fusion outperforms condorcet and individual rank learning methods},
  author={Cormack, Gordon V and Clarke, Charles LA and Buettcher, Stefan},
  booktitle={Proceedings of the 32nd international ACM SIGIR conference on Research and development in information retrieval},
  pages={758--759},
  year={2009}
}

@article{robertson2009probabilistic,
  title={The probabilistic relevance framework: BM25 and beyond},
  author={Robertson, Stephen and Zaragoza, Hugo and others},
  journal={Foundations and Trends{\textregistered} in Information Retrieval},
  volume={3},
  number={4},
  pages={333--389},
  year={2009},
  publisher={Now Publishers, Inc.}
}

@article{wang2022text,
  title={Text embeddings by weakly-supervised contrastive pre-training},
  author={Wang, Liang and Yang, Nan and Huang, Xiaolong and Jiao, Binxing and Yang, Linjun and Jiang, Daxin and Majumder, Rangan and Wei, Furu},
  journal={arXiv preprint arXiv:2212.03533},
  year={2022}
}

@inproceedings{glass2022re2g,
  title={Re2G: Retrieve, Rerank, Generate},
  author={Glass, Michael and Rossiello, Gaetano and Chowdhury, Md Faisal Mahbub and Naik, Ankita and Cai, Pengshan and Gliozzo, Alfio},
  booktitle={Proceedings of the 2022 Conference of the North American Chapter of the Association for Computational Linguistics: Human Language Technologies},
  pages={2701--2715},
  year={2022}
}

@article{yu2024rankrag,
  title={Rankrag: Unifying context ranking with retrieval-augmented generation in llms},
  author={Yu, Yue and Ping, Wei and Liu, Zihan and Wang, Boxin and You, Jiaxuan and Zhang, Chao and Shoeybi, Mohammad and Catanzaro, Bryan},
  journal={Advances in Neural Information Processing Systems},
  volume={37},
  pages={121156--121184},
  year={2024}
}

@article{moreira2024enhancing,
  title={Enhancing Q\&A Text Retrieval with Ranking Models: Benchmarking, fine-tuning and deploying Rerankers for RAG},
  author={Moreira, Gabriel de Souza P and Ak, Ronay and Schifferer, Benedikt and Xu, Mengyao and Osmulski, Radek and Oldridge, Even},
  journal={arXiv preprint arXiv:2409.07691},
  year={2024}
}

@article{freja2026evalqreason,
  title={EvalQReason: A Framework for Step-Level Reasoning Evaluation in Large Language Models},
  author={Freja, Shaima Ahmad and Catak, Ferhat Ozgur and Yurdem, Betul and Rong, Chunming},
  journal={arXiv preprint arXiv:2602.02295},
  year={2026}
}

@article{joshi2017triviaqa,
  title={Triviaqa: A large scale distantly supervised challenge dataset for reading comprehension},
  author={Joshi, Mandar and Choi, Eunsol and Weld, Daniel S and Zettlemoyer, Luke},
  journal={arXiv preprint arXiv:1705.03551},
  year={2017}
}

@inproceedings{rabinovich2023predicting,
  title={Predicting question-answering performance of large language models through semantic consistency},
  author={Rabinovich, Ella and Ackerman, Samuel and Raz, Orna and Farchi, Eitan and Tavor, Ateret Anaby},
  booktitle={Proceedings of the Third Workshop on Natural Language Generation, Evaluation, and Metrics (GEM)},
  pages={138--154},
  year={2023}
}

@article{schulhoff2024prompt,
  title={The prompt report: a systematic survey of prompt engineering techniques},
  author={Schulhoff, Sander and Ilie, Michael and Balepur, Nishant and Kahadze, Konstantine and Liu, Amanda and Si, Chenglei and Li, Yinheng and Gupta, Aayush and Han, HyoJung and Schulhoff, Sevien and others},
  journal={arXiv preprint arXiv:2406.06608},
  year={2024}
}

@article{laskar2024systematic,
  title={A systematic survey and critical review on evaluating large language models: Challenges, limitations, and recommendations},
  author={Laskar, Md Tahmid Rahman and Alqahtani, Sawsan and Bari, M Saiful and Rahman, Mizanur and Khan, Mohammad Abdullah Matin and Khan, Haidar and Jahan, Israt and Bhuiyan, Amran and Tan, Chee Wei and Parvez, Md Rizwan and others},
  journal={arXiv preprint arXiv:2407.04069},
  year={2024}
}

@article{nguyen2016ms,
  title={Ms marco: A human-generated machine reading comprehension dataset},
  author={Nguyen, Tri and Rosenberg, Mir and Song, Xia and Gao, Jianfeng and Tiwary, Saurabh and Majumder, Rangan and Deng, Li},
  year={2016}
}

@inproceedings{razavi2025benchmarking,
  title={Benchmarking prompt sensitivity in large language models},
  author={Razavi, Amirhossein and Soltangheis, Mina and Arabzadeh, Negar and Salamat, Sara and Zihayat, Morteza and Bagheri, Ebrahim},
  booktitle={European Conference on Information Retrieval},
  pages={303--313},
  year={2025},
  organization={Springer}
}


\end{document}